# Method of problem solution of diffraction and scattering theory

*Morozov Valery B.*

*Problem solutions in area of diffraction and of scattering theory are considered from one point of view. The method common for them is based on approximate orthogonality of solution constituents, which oscillate on a body long frontier. Method potentiality is discussed.*

As an original equation, the following integral equation can be taken:

$$A u_0(\mathbf{r}) = \int v(\xi) A D(\mathbf{r},\xi) \, d\xi, \qquad (1)$$

where $A u(\mathbf{r}) = 0, \ \mathbf{r} \in S$ is a boundary condition and $A$ is a linear operator.

The problem solution is represented in the following form:

$$u(\mathbf{r}) = u_0(\mathbf{r}) + \int v(\xi) D(\mathbf{r},\xi) \, d\xi, \qquad (2)$$

where $u_0(\mathbf{r})$ is nonperturbed wave field.

If the problem solution is represented in a form of countable set of sum members (degenerated case), it is necessary to replace integrals having integrating parameter $\xi$ with equivalent sums.

After substituting of the expression

$$v(\xi) = \int_S \beta [A D(\mathbf{r}',\xi)]^* w(\mathbf{r}') \, d\mathbf{r}' \qquad (3)$$

and an integration order change, we can transform the expression (1):

$$A u_0(\mathbf{r}) = \int_S w(\mathbf{r}') \Phi(\mathbf{r},\mathbf{r}') \, d\mathbf{r}', \qquad (4)$$

where the nucleus is

$$\Phi(\mathbf{r},\mathbf{r}') = \int \beta [A D(\mathbf{r}',\xi)]^* A D(\mathbf{r},\xi) \, d\xi, \qquad (5a)$$

notably $\beta$ is normalizing factor, which in general depends on $\xi$.

Or in the degenerated case

$$\Phi(\mathbf{r},\mathbf{r}') = \sum_i \beta_i [A D_i(\mathbf{r}')]^* A D_i(\mathbf{r}), \qquad (5b)$$

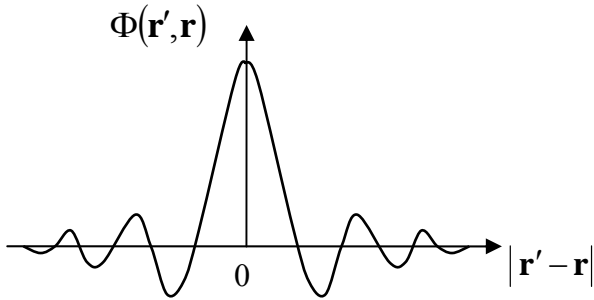

Fig. 1. Typical functional connection of the nucleus $\Phi(\mathbf{r'},\mathbf{r})$ versus the module $|\mathbf{r'}-\mathbf{r}|$

If under prescribed functions $D(\mathbf{r},\xi)$ in a problem a variables separation is allowed and functions $AD(\mathbf{r},\xi)$ are orthogonal on S, then in a case of corresponding normalization the nucleus (5) transforms into $\delta$-function. As this takes place, the equation (4) will take the following form:

$$w(\mathbf{r'}) = -A\beta u_0(\mathbf{r}), \qquad (6)$$

which in this case with taking (3) into account, leads to exact solution of the diffraction problem.

But if $\Phi(\mathbf{r'},\mathbf{r})$ decays quickly enough in proportion to $|\mathbf{r'}-\mathbf{r}|$ growth (Fig. 1), one may say about approximate orthogonality of the function set $AD(\mathbf{r},\xi)$ or the function set $AD_i(\mathbf{r})$. This feature of the nucleus is peculiarly pronounced for speedily oscillating functions, i.e. for solution of undular problems in a case of high enough values of ratio of a typical size scatterer versus a wave length. Under these stipulations, the equation (6) yields the approximation:

$$w_1(\mathbf{r'}) = -A\beta u_0(\mathbf{r}), \qquad (7)$$

which normalizes the factor

$$\beta = \left[\int d\mathbf{r'}\int [AD(\mathbf{r'},\xi)]^* AD(\mathbf{r},\xi) d\xi\right]^{-1}$$

Here as a small parameter, the value

$$\varepsilon = \frac{\int [AD(\mathbf{r'},\chi)]^* AD(\mathbf{r},\xi) dr}{\int [AD(\mathbf{r'},\xi)]^* AD(\mathbf{r},\xi) dr}|v(\chi)-v(\xi)|$$

can arise.

From (4) and (7), the interrelation

$$w_1(\mathbf{r}) = \beta \int_S w(\mathbf{r'})\Phi(\mathbf{r},\mathbf{r'}) d\mathbf{r'}, \qquad (8)$$

can be received connecting an exact value $w(\mathbf{r})$ with approximated one $w_1(\mathbf{r})$. This expression allows considering of the approximated solution $w_1(\mathbf{r})$ as a however leveled accurate solution, transforming into an exact one on approach to the limit $\Phi(\mathbf{r},\mathbf{r'}) \to \delta(\mathbf{r}-\mathbf{r'})$.

The problem (1) belongs to incorrect problems [1]. The interrelation (8) can be considered as a functional of solution regularization. This makes it possible to take up the position that iterations on the proposed method are going to be stable and converging.

Use of the solution (2) represented in a form of superposition of Green's functions $G(\mathbf{r}',\mathbf{r})$ and boundary conditions $A u(\mathbf{r}) = 0$ with functional $A = \nabla^2 + k^2 + \Xi(\mathbf{r})$ enables solving of a problem of field scattering in a non-homogeneous medium. The method concerned leads to the solution:

$$u = u_0 - \int \beta(\mathbf{r}')G(\mathbf{r}',\mathbf{r})\Xi(\mathbf{r}')u_0(\mathbf{r}')d\mathbf{r}' - \iint \beta(\mathbf{r}')G(\mathbf{r}'',\mathbf{r}')G(\mathbf{r}',\mathbf{r})|\Xi(\mathbf{r}'')|^2 u_0(\mathbf{r}')d\mathbf{r}'d\mathbf{r}''.$$

The expression received puts in mind of Born's second approximation. However the second integral depends only on absolute disturbance value $\Xi$; In both integrals, a normalizing factor is additionally present, while the second integral sign is opposite to the sign of the corresponding integral of the Born's approximation.

An oscillating nature of the functions $A D(\mathbf{r},\xi)$ gives a reason to consider the parameter $\varepsilon$ being small and staying small on a long enough integration domain *S*, but along with it a surface *S* form can be complicated enough.

This assumption can be stated in a form of a theorem showing that if $\chi \neq \xi$ and a modulus of a wave number $k \to \infty$, then

$$\int [A D(\mathbf{r}',\xi)]^* A D(\mathbf{r},\xi) d\xi \to 0.$$

If to represent the solution in a form of plane waves, this theorem coincides with multidimensional version of Riemannian theorem on limit of integral, which contains a harmonic function.

One more problem, on which the method was verified, was a diffraction problem of a plain acoustic wave on an absolutely hard sphere. The problem solution was represented in the form of the superposition of the plain waves, to which attention was paid in the work [2]; And, indeed, the solution has a zero amplitude of an imaginary component of a reflection coefficient. However the reflection coefficient dependences received with use of the proposed method reproduce well features of the well-known exact solution for both acoustically hard sphere and acoustically soft one.

For plain screens featured with an acoustically hard surface and a solution containing a plain waves representation (naturally in this case also not complete one), the method concerned coincides with accurateness to constant factor with Kirchhoff's method. Such result can be considered as the vindication of Kirchhoff's heuristic approach. However unlike Kirchhoff's approach, our method is applicable also to the non-plain screens and any boundary conditions on the screen surface. Numerical calculations for the slit diffraction problem showed that the method at issue reproduces features of the diffraction field exact solution [3] not only for acoustically hard screens, but also for acoustically soft ones.

As an interesting thing, there proved to be testing results of the method with use of a solution representation as a combination of a sources finite number. The problem of plain wave diffraction was solved on a hard prolate spheroid; This problem has got the rigorous solution [4]. The problem solution was searched in a form of a sum of fields of several sources located on spheroid axis: The farfield method has yielded satisfactory outcomes for orthogonalization on surface.

Previously the similar approach was proposed for an approximate separation of wave equation variables in boundary problems [5].

Author hopes that the method will prove to be useful for solving or for solutions qualitative evaluation for problems of diffraction and quantum mechanics. Author is appreciative to Lev Albertovich Veinstein of his interest to this work.

*References*

1. Tikhonov A.N., Arsenin V.Ya. Solution methods of incorrect problems. Moscow: Science, 1986.–288 pages. (In Russian).
2. Veinstein L.A., Suekov A.I. Diffraction on surface with periodic relief (undulating surface). Preprint #8(380) IRE of Academy of Sciences of USSR. 1984. (In Russian).
3. Hönl H., Maue A.W., Westpfahl K. Theorie der Beugung. Springer. 1961.
4. Konyukhova N.B, Pak T.V. Plain wave diffraction of hard prolate spheroid. Reports on applied mathematics. Moscow: Computer Center of Academy of Sciences of USSR. 1985. (In Russian).
5. Veinstein L.A. Method of approximated variables separation and its application to boundary problems of electrodynamics and acoustics. – Technical physics journal, 1957, volume 27, # 9, pages 2109-2128. (In Russian).